\documentclass[aps,superscriptaddress,twocolumn,showpacs,preprintnumbers,amsmath,amssymb]{revtex4}
\usepackage[cp1251]{inputenc}
\usepackage{amssymb,amsmath}
\usepackage[mathscr]{eucal}
\usepackage{graphicx}
\usepackage{longtable}
\usepackage[dvips]{hyperref} 
\begin{document}
\title{Rabi Waves in Carbon Nanotubes - Experiment}
\author{Dmitry Yearchuck (a), Alla Dovlatova (b)\\
\textit{(a) - Minsk State Higher Aviation College, Uborevich Str., 77, Minsk, 220096, RB; yearchuck@gmail.com; \\ (b) - M.V.Lomonosov Moscow State University, Moscow, 119899}}
\date{\today}
\begin{abstract} Rabi waves have been experimentally registered for the first time by Raman scattering studies of zigzag nanotubes, produced by high energy ion beam modification of natural diamond single crystals. Antiferroelectric spin wave resonance has been detected for the first time in Raman spectroscopy practice in given samples. Substantial qualitative and quantitative changes in Raman spectra in dependence on propagation direction of laser excitation wave have been found.
\end{abstract}
\maketitle 
The relativistic theory of quantized fields, in particular, quantum electrodynamics (QED) takes on more and more significance for its practical application and it, in fact, becomes to be working instrument in spectroscopy studies and industrial spectroscopy control. Presented work confirms given conclusion.
The authors  of recent publications \cite{Slepyan}, \cite {Slepyan_Yerchak}, in which  the generalization  of multiqubit QED-model of  \cite {Tavis} has been proposed by taking into account the interaction between the qubits and by inclusion in Hamiltonial of  nonlinear term, have also predicted and theoretically studied the new coherent effect of nonlinear quantum optics -- spatial propagation of Rabi oscillations - Rabi waves - in one-dimensional (1D) quantum dot  chain. The similar effect is predicted in \cite{Part1} for interacting multichain system, on the example of zigzag nanotubes (NT), in which the chains, being to be connected each other, are naturally strongly interact between themselves. 
The aim of given letter is experimental confirmation of the conlusions of the works \cite{Slepyan}, \cite {Slepyan_Yerchak}, \cite{Part1}, that is experimental evidence for the existence of  Rabi wave phenomenon.
Samples of type IIa natural diamond (nitrogen content was less than $5\times{10^{17}}$ $cm^{-3}$), implanted by high energy ions of copper $(63$ $MeV$, $5\times{10^{14}}$ $cm^{-2}$) and boron $(13,6$ $MeV)$ have been studied. Ion implantation was performed along $\left\langle{111}\right\rangle$ crystal direction. Raman scattering  (RS) spectra were registered in backscattering geometry. Laser excitation wave length was 488 $nm$, rectangular slit $ 350{\times}350 (\mu m)^2$ was used, scan velocity was 100 $cm^{-1}$ pro minute. The spectra observed are presented in Figures 1 to 3. We see from comparison of the spectra, presented in Figures 1 and 2, that they are strongly dependent on laser beam direction. The spectrum, presented in Figure 1,  is characteristic of the only ion beam modified region of the sample, since characteristic  diamond line near 1332 $cm^{-1}$ is absent in the spectrum. The RS-lines with peak positions $656.8{\pm}0.2$ $cm^{-1}$, $1215{\pm}1$ $cm^{-1}$, $1779.5{\pm}1$ $cm^{-1}$ and  $2022.3{\pm}0.5$ $cm^{-1}$ correspond to given region by laser excitation transversely to sample surface from implanted side.  At the same time by laser excitation of the same sample from opposite unimplanted side the qualitatively other picture is observed, see Figure 2. Firstly, rather intensive relatively narrow line, which is characteristic for diamond single crystals, is now naturally presented in the spectrum (full amplitude of given line is not shown in Figure 2). It is interesting, that its frequency value, equaled to 1328.7 $cm^{-1}$ is slightly shifted from usually observed value near 1332 $cm^{-1}$. It demonstrates the implantation effect on the whole sample, indicating on renormalization of optical phonon in diamond matrix, although the  thickness of the sample  $\simeq 1 mm $ is much greater, than the thickness  of the near-surface ion beam modified region, which does not excced $7 \mu m$, (implanted copper atom concentration is characterized by narrow peak with maximum at $\approx 6.5 \mu m$) \cite {Erchak}. Secondly, the spectrum of the same ion beam modified region is characterized by the quite different set of lines. The RS-lines with peak positions 165.3 (feature), 209.8 (feature), the lines at 354.6, 641.8, 977.1 (${\pm}1$ $cm^{-1}$), 1274.1 ${\pm}2$ $cm^{-1}$  and weakly pronounced lines at 1569 ${\pm}3$ $cm^{-1}$, 1757${\pm}5$ $cm^{-1}$ were observed in RS-spectrum, corresponding to given case. Moreover, all the lines listed are superimposed now with very broad (its linewidth value is 1720${\pm}20$ $cm^{-1}$) asymmetric line with peak position $1160{\pm}10$$cm^{-1}$. 
\begin{figure}
\includegraphics[width=0.5\textwidth]{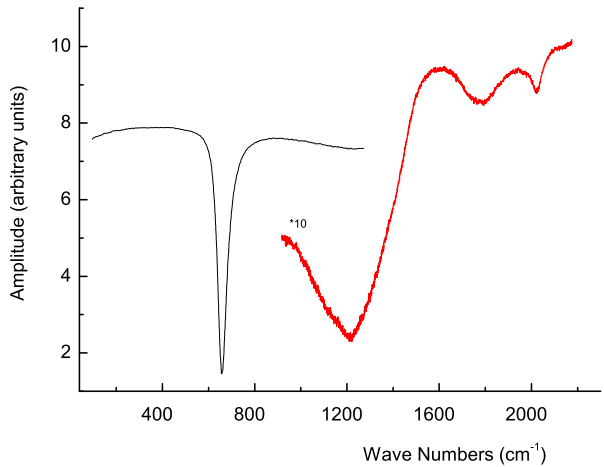}
\caption[Spectral distribution of Raman scattering  intensity in
diamond single crystal, implanted by high energy copper ions, the excitation is from implanted side of the sample.]
{\label{Figure1} Spectral distribution of Raman scattering  intensity in
diamond single crystal, implanted by high energy copper ions, the excitation is from implanted side of the sample.}
\end{figure}
\begin{figure}
\includegraphics[width=0.5\textwidth]{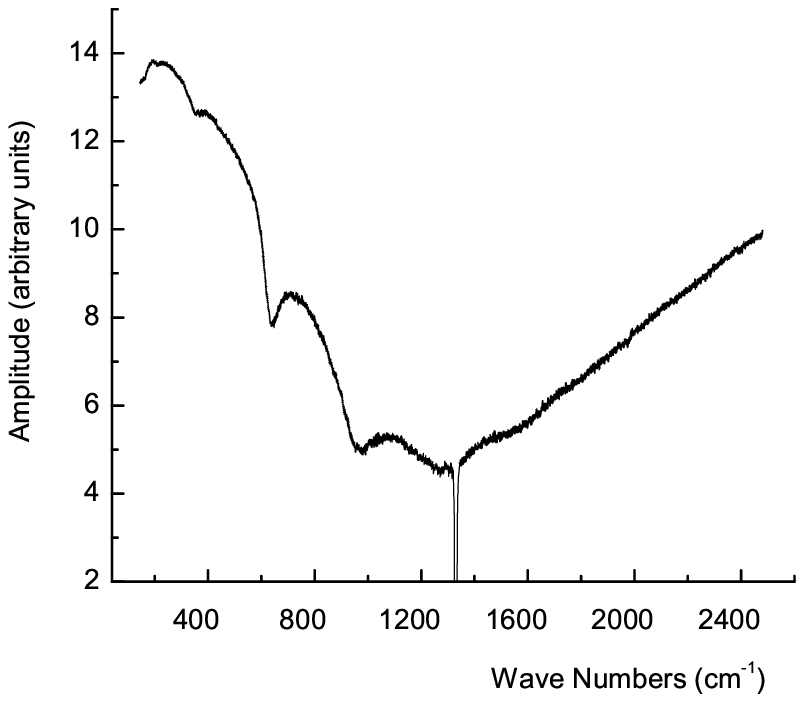}
\caption[Spectral distribution of Raman scattering  intensity in
diamond single crystal, implanted by high energy copper ions, the excitation is from unimplanted side of the sample.]
{\label{Figure2} Spectral distribution of Raman scattering  intensity in
diamond single crystal, implanted by high energy copper ions, the excitation is from unimplanted side of the sample.}
\end{figure}
\begin{figure}
\includegraphics[width=0.5\textwidth]{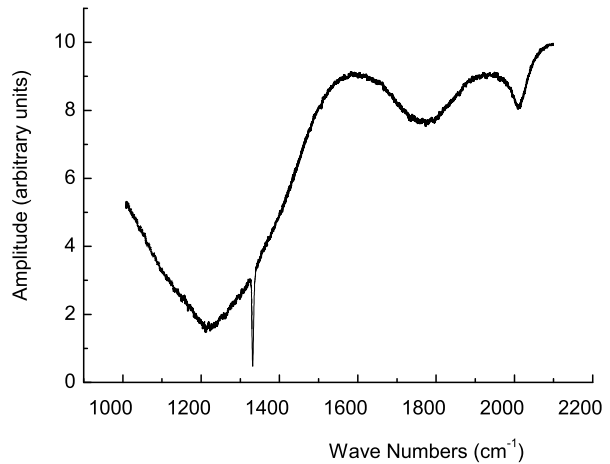}
\caption[Spectral distribution of Raman scattering  intensity in
diamond single crystal, implanted by high energy boron ions, the excitation is from implanted side of the sample.]
{\label{Figure3} Spectral distribution of Raman scattering  intensity in
diamond single crystal, implanted by high energy boron ions, the excitation is from implanted side of the sample.}
\end{figure}
The spectrum of boron implanted sample was measured in the range (1000 - 2100) $cm^{-1}$ by excitation from implanted side. It is seen, that the spectra of boron and copper implanted samples are qualitatively similar in given spectral range. However numerical values of the peak positions are slightly different, they are - 1212.3${\pm}1$,  1772.5${\pm}1$, 2011${\pm}0.5$$cm^{-1}$ in the spectrum, which is characteristic of the only ion beam modified region and characteristic diamond line with peak position 1331.95${\pm}0.1$$cm^{-1}$ was also presenting. Its frequency value coincides with the value of optical phonon peak in conventional natural diamonds. The presence of diamond RS-line by excitation from implanted side can be determined by two factors. Firstly the effective thickness of ion beam modified region is substantially less $(\simeq{1.4}\mu m)$, it allows to suggest, that laser excitation can reach an unimplanted region. It can also suggest, that the modification is not entire in near surface region. It is interesting, that there is regularity in peak position shift, which is increasing  with frequency increase and it is equal to 2.8, 7, 11.3 $cm^{-1}$ correspondingly. It is strong indication, that all three lines belong to the same optical system. 
To discuss the results of RS-studies, let us represent the model of structures, produced by high energy ion implantation. 
A number of the  studies with various methods and with the samples implanted with various ions, energy, fluences has been performed. See, for instance, review article \cite{ErtchakD} on ESR-studies, which were the most informative in structure identification. It was established the formation of quasionedimensional spatially ordered tracklike carbon structures, that is the structures with diameter of nanorange. It was in fact the discovery of new carbon phase - carbon nanotubes, incorporated in diamond matrix, or occupying all near surface volume region. It is remarkable, that the first report is related to 1990, which was made  during 1990 IBMM-Conference, Knoxwille, USA and it was repeated at E-MRS 1990 Fall Meeting, Strasbourg, France, that is nothing was at that time known on the japan discovery of free NTs, related to 1991. All subsecuent studies have confirmed given initial conclusion \cite{ErtchakD}. The most designing among the results obtained were the values of $g$-tensor components. They are in $Cu$-implanted sample $g_1$ = 2.00255 (it is minimal $g$-value and it is $g_{||}$ principal direction of axial g-tensor, at that  it  coincides with ion beam direction), $g_2 = g_3$= $g_\perp$ = 2.00273, the accuracy of relative $g$-value measurements is $\pm 0.00002$ \cite{Erchak}. So we see on the one hand, that it was possible to determine anisotropy of $g$-values with very high precision, that indicates on the very perfect and homogeneous NTs, produced with strict axial symmetry along ion beam direction, if to take into account the origin of  paramagnetic centers (PC) in ion produced NT  to be  paramagnetic $\pi$-solitons   (SSH-solitons), considered to be mobile PC, mapping the distribution of $\pi$-electron density along whole individual chain \cite{ErtchakD}, \cite{Ertchak}.
On the other hand,  the g-value of paramagnetic $\pi$-solitons in trans-polyacetylene, equaled to  
2.00263 \cite{Goldberg}, gets in the middle of given rather narrow interval of g-value variation of  PC in ion produced NT. Although anisotropy of paramagnetic $\pi$-solitons in t-PA, which are also considered to be mobile paramagnetic centers, mapping the distribution of $\pi$-electron density along whole individual t-PA chain, is not resolved by ESR measurements directly (which in fact is the indication, that chemically produced t-PA is less perfect in comparison with NTs in diamond matrix), there are indirect evidences on axial symmetry of $\pi$-solitons in t-PA too, \cite{Kahol} \cite{Kuroda}. Consequently, the value 2.00263 is mean value and it coincides with accuracy 0.00002 with mean value of aforecited principal g-tensor values  of PC in NTs. Given coincidence becomes to be understandable now, if to take into account the results of \cite{Part1}, where is shown, that SSH-model is applicable to zigzag NT.  In the frames of SSH-model  zigzag NT is considered to be the system of $n$ equivalent t-PA  chains interacting between themselves. 
Let us give some details, concerning SSH-model.
It is remarkable, that SSH-model contains in implicit form along with the physical basis for the existence  of solitons, polarons, breathers, formed in $\pi$-electronic subsystem ($\pi$-solitons, $\pi$-polarons, $\pi$-breathers), also the basis for the existence of similar quasiparticles in  $\sigma$-electronic subsystem, that is $\sigma$-solitons, $\sigma$-polarons, $\sigma$-breathers. The cause  is the same two-fold degeneration of ground state  of the whole electronic system, energy of which has the form of Coleman-Weinberg potential with two minima at the values of dimerization coordinate $u_0$ and $ -u_0$. 
The shapes, for instance, of $\pi$-solitons and
$\sigma$-solitons can be given by the expression with the same mathematical form
\begin{equation}
\label{Eq1}
|\phi(n)|^2 = \frac{1}{\xi_{\pi(\sigma)}} sech^2[\frac{(n-n_0)a}{\xi_{\pi(\sigma)}} - v_{\pi(\sigma)} t] cos \frac{n \pi}{2},
\end{equation} 
where $n, n_0$ are variable and fixed numbers of $CH$-unit in $CH$-chain, $a$ is $C-C$ interatomic spacing projection on chain direction, $v_{\pi(\sigma)}$ is $\pi$($\sigma$)-soliton velocity, $t$ is time, $\xi_{\pi(\sigma)}$ is $\pi$($\sigma$) coherence length. It is seen, that $\pi$-solitons and $\sigma$-solitons differs in fact the only by numerical value of coherence length. Given difference can be evaluated even without numerical calculation of the relation, which determines the shift of ground state energy of extended system by presence of localized perturbation. Actually it is sufficient  to take into account the known value of $\xi_\pi$ and relationships \cite{Lifshitz}
\begin{equation}
\label{Eq2}
\xi_{0\pi} = \frac{\hbar v_F}{\Delta_{0\pi}}, \xi_{0\sigma} = \frac{\hbar v_F}{\Delta_{0\sigma}},
\end{equation}
where $\Delta_{0\sigma}$, $\Delta_{0\pi}$ are $\sigma-$ and $\pi$-bandgap values at $T = 0 K$, $v_F$ is Fermi velocity.
Theoretical value $\xi_\pi$ in t-PA is $7a$, and it is low boundary in the range 
$7a - 11a$, obtained for $\xi_\pi$ from experiments \cite{Heeger_1988}. Taking into account the relationships (\ref{Eq2}),  using the value $\frac{\Delta_{\sigma}}{\Delta_{\pi}}\approx 8.8$, which was evaluated from t-PA band structure calculation in \cite{Grant}, and mean experimental value of coherence length $\overline{\xi_{\pi}} = 9a$ we obtain the value $\overline{\xi_{\sigma}} \approx 0.125 nm$. It means, that the width of space region, occupied by $\sigma$-soliton in t-PA is $\approx 0,5 nm$, that is SSH-$\sigma$-solitons are much more localized in comparison with SSH-$\pi$-solitons. Similar conclusion takes place for SSH-$\sigma$-polarons representing itself the soliton-antisoliton pair, that is the width of space region, occupied by SSH-$\sigma$-polarons is evaluated to be $\approx 1 nm$.   SSH-$\sigma$-polarons have recently been experimentally detected in related material -carbynes \cite{Yearchuck_PL}, where the formation of polaron lattice (PL) was proposed.
 It was established, that  two components of each elementary unit, that is, of each polaron, possess by two equal in values electical own dipole  moments, proportional to spin, which was called electical spin moments (ESM), with opposite directions. It was shown, that experimental results agree well with  proposal of PL-formation, which means in fact  
 the formation of antiferroelectrically ordered lattice of quasiparticles. Given lattice consists of 2 sublattices, corresponding to soliton and antisoliton components of polaron. Corresponding chain state is optically active and it is characterized by  the set of lines in IR-spectra, which were assigned with new phenomenon - antiferroelectric spin wave resonance (AFESWR)
Central mode is convential antiferroelectric  resonance (AFR)  mode, its value $\nu^\sigma_p(C)$ in carbyne sample studied was 477 $cm^{-1}$. Let us remember that carbynes are organic quasionedimensional conductors with the simplest, consisting the only of the carbon atoms, chain structure. At the same time the presence of two electronic  $\pi_[$- and $\pi_y$-subsystems, which are "hung" on $\sigma$-subsystem means thatthe ground electronic state is similar to twodimensional Coleman-Weinberg potential with four minima at the values of dimerization coordinate $u_0$ and $ -u_0$. In other words, ground electronic state in carbynes is four-fold degenerate, which leads to a substantially more rich spectrum of possible quasiparticles, discussed in \cite{Rice}, \cite{Yearchuck_ArXiv}.

Known value $ \nu^\sigma_p(C)$ of the frequency of SSH-$\sigma$-polaron IR-active mode in carbynes allows to estimate the range for expected values 
$\nu^\sigma_p(t-PA)$ and $\nu^\sigma_p(NT)$ for the frequencies of SSH-$\sigma$-polaron IR-active mode in t-PA and in nanotubes of zigzag kind. Really, the known relationship for the vibration frequencies of similar centers, that is the relationship
\begin{equation}
\label{Eq3}
\frac{\nu^{\sigma}_p(t-PA)}{\nu^{\sigma}_p(C)} = 
\sqrt{\frac{g(C) m^{\sigma}_p(C)}{g(t-PA) m^{\sigma}_p(t-PA)}}
\end{equation}
is always takes place. Here $m^{\sigma}_p(C)$
$m^{\sigma}_p(t-PA)$ are SSH-$\sigma$-polaron masses, $g(C)$, $g(t-PA)$ are degeneration degree factors in carbynes and in t-PA respectively.  It is evident, that the ratio
of polaron masses can be replaced by the soliton mass ratio, the expression for which is \cite{Su_Schrieffer_Heeger_1980}, \cite{Heeger_1988}
\begin{equation}
\label{Eq4}
m_s = \frac{4}{3\xi}( \frac{u_0}{a})^2 M, 
\end{equation}
where $M$ is mass of $C-H$ unit and carbon atom in the case of t-PA and carbyne respectively. Then, taking into account the relationships (\ref{Eq2}) we obtain
$\nu^\sigma_p(t-PA) \in$ (487.2, 759.7) $cm^{-1}$ and $\nu^\sigma_p(NT) \in $(507.1, 790.7) $cm^{-1}$. The theoretical values for carbyne $\sigma$-bandgap from calculation in \cite{Leleiter} were used. They are strongly dependent on the number of carbon atoms in elementary units of carbyne chain, that determines the relatively large spectral regions above indicated. Let us compare the expected value for IR-active SSH-$\sigma$-polaron mode in t-PA with experimental data. The mode with the frequency near 540 $cm^{-1}$, which was observed in infrared photoinduced   spectra of t-PA gets to interval (487.2, 759.7) $cm^{-1}$ and can represent itself AFR  mode in $\sigma$-polaron lattice. It means, that earlier ascribing IR-line near 540 $cm^{-1}$ to Goldstone SSH-$\pi$-soliton mode has to be reinterpreted. Given conclusion is confirmed additionally by the following arguments.
1.The frequency value in $\sim 540$ $cm^{-1}$ is too large for any Goldstone mode. 
For comparison, so called "breathing" modes  in related materials (see also further consideration) - carbon single wall NTs - were observed  in the range between 140-200 $cm^{-1}$ \cite{Anglaret}, moreover shear modes  in nested NTs have the frequency values at 49, 58 $cm^{-1}$ \cite{Eklund}. Similar values  of the frequency seems to be reasonable to expect for Goldstone mode of SSH-$\pi$-solitons. It is remarkable, that at the same time  the lines at $\sim 700 cm^{-1}$ \cite{Eklund} in nested NTs and in the range 750-790 $cm^{-1}$ for single wall NTs \cite{Anglaret} were also presenting. Naturally,  authors of \cite{Anglaret}, \cite{Eklund} do not refer given modes to Goldstone modes, although their frequency values are intermediate between $\sim 540$ $cm^{-1}$ and $\sim 900$ $cm^{-1}$, which were considered to be Goldstone modes in undoped and doped t-PA respectively. It is evident, that the pinning role in undoped NT and undoped  t-PA is comparable.
2.The resemblance in the set of lines, which belong to the same absorbing center and which are  in given case the result of fine (vibronic) splitting of the same electronic level has to be revealed. The  resemblance is concerned first of all
 the line shapes and linewidths. Comparison of line shapes and linewidths of the line $\sim 540 cm^{-1}$ and the second line with higher frequency value in Fig.39 \cite{Heeger_1988} shows, that the differences in line shapes and linewidths  are very large. It means, that compared lines cannot belong to the same center.
3.Both the lines $\sim 540$ $cm^{-1}$ and $\sim 900$ $cm^{-1}$  are strongly intensive. It testifies to the favour of proposal, that $\sim 540$ $cm^{-1}$ absorption line  in photoexcited t-PA is  analogue of 477 $cm^{-1}$ carbyne line, that is, it can really be assigned  with AFR  mode in $\sigma$-polaron lattice, produced in  t-PA by photoextitation along with charged $\pi$-soliton formation. We also wish to remark, that the line $\sim 900$ $cm^{-1}$ can also belong to $\sigma$-polarons in another charge state and that AFSWR can in principle  be observable by suitable geometry of experiment in both the cases.  

Taking into account the calculation of the frequecies of vibration active modes for NTs of various kind in \cite{Eklund}, we can conclude a priori, without additional calculations, that in spectral interval, which is near to (507.1, 790.7) $cm^{-1}$ $\sigma$-polaron in zigzag NT will have the vibration active mode, which is active in Raman scattering. Really, on the one hand, the formation of quasiparticles in zigzag NT of the same type, that those ones in t-PA, was proved in \cite{D_Y}. On the other hand, the calculation  in \cite{Eklund} indicates on the presence of even and uneven modes in above indicated interval with frequencies, which have close values, at that some of even Raman active modes are dependent on NT diameter value. Given calculation does not take into consideration the soliton and polaron formation. However $\sigma$-polaron or $\sigma$-soliton formation does not violate the symmetry of task, that means, that for zigzag NT to the IR $\sigma$-polaron mode in range (507.1, 790.7) $cm^{-1}$  will correspond the Raman $\sigma$-polaron mode  at close frequency range.
 Therefore we come to conclusion, that intensive line $656.8{\pm}0.2$ $cm^{-1}$ in RS-spectrum, presented in Fig.1 can be assigned with AFR mode of $\sigma$-polaron lattice, produced in NTs, while the lines, $1215{\pm}1$ $cm^{-1}$, $1779.5{\pm}1$ $cm^{-1}$ and  $2022.3{\pm}0.5$ $cm^{-1}$ is revival part\cite {Slepyan_Yerchak}  in its frequency representation \cite{Part1} of Rabi wave packet, which is produced in result of interaction of $\sigma$-polaron lattice with quantized EM-field. Given identification is confirmed by the following. It is well known, that is in the case of point absorbing centers Rabi frequency is linearly dependent on the amplitude of oscillating EM-field. To lesser value of exiting amplitude correspond also lesser values of the frequences of  traveling Rabi waves. Given tendency takes also place for high frequency components of wave packets, that follows from the analysis of Fourier transform of temporal dependence  of the  integral inversion for a coherent initial state of light, given in \cite {Slepyan_Yerchak}. Further, it is evident, that amplitude of laser wave, penetrating in ion beam modified region, is less by excitation from unimplanted side in consequence of some absorption in the unimplanted volume of diamond single crystal. We see, that really, the experimental frequency values 1569 ${\pm}3$ $cm^{-1}$, 1757${\pm}5$ $cm^{-1}$ of two high frequency components are substantially lesser in given case, than $1779.5{\pm}1$ $cm^{-1}$ and  $2022.3{\pm}0.5$ $cm^{-1}$, observed by excitation from implanted side, compare Fig.2 and Fig.1. The  AFR  mode is observed now at 641.8 ${\pm}1$ $cm^{-1}$ instead of  $656.8{\pm}0.2$ $cm^{-1}$, which indicates on the presence of hysteresis relatively the direction of excitation wave propagation. The lines at 354.6,  977.1 (${\pm}1$ $cm^{-1}$ seem to be assigned with two AFSWR 
modes, that is, there is splitting of AFR-mode into AFSWR spectrum. The substantial decrease of relative intensity of 641.8 ${\pm}1$ $cm^{-1}$ mode in comparison with  $656.8{\pm}0.2$ $cm^{-1}$ testifies in favour of given assignment. It is seen, that AFSWR-splitting is rather large and it is comparable with the splitting between two polaron vibration levels. It means, that linear AFSWR-theory, developed in \cite{Yearchuck_PL}, which predicts   a set of equidistant  AFSWR-modes, arranged the left and the right of central mode, will be not applicable for given case. Really  AFSWR-modes  are not equidistant, they are shifted on distances 335.3, 287,2 $cm^{-1}$ from main AFSWR-mode. At the same time mean value of AFSWR-splitting is 311.3$cm^{-1}$ and it is close to two-fold value of  AFSWR splitting (which is equal to 150 $cm^{-1}$ assigned with $\sigma$-polaron lattice in carbynes \cite{Yearchuck_PL}. Given approximately two-fold splitting increase seems to be result of corresponding increase of  spin wave resonance splitting, registered by Raman spectroscopy methods in comparison with SWR splitting, registered by IR spectroscopy methods, predicted in \cite{Yearchuck_Doklady} and observed earlier by ferroelectric SWR study in carbynes \cite{Yearchuck_Yerchak}. To explain so significant differences in the physics of processes by the change of the direction of the excitation wave propagation, we have to take into consideration the following. Any SWR-excitation in 1D-systems is strongly dependent on the geometry of experiment, that is, on the mutual  orientation of chain axis and vectors of intracrystal electric field, which determines the splitting of energetic levels, and electrical component of EM-field, calling the spectroskopic transitions like to possibility of the excitation of ferromagnetic spin wave resonance in carbynes \cite{Ertchak_J_Physics_Condensed_Matter}. NT axes in zigzag NTs are not linear in the end of ion run and they are generatrixes  of the figure of onion-like shape, that provides for necessary geometry for  AFSWR excitation in photon dressed state. Moreover the appearance of very broad line seems to be indication on the excitation by given laser wave propagation  direction of the movement of polaron lattice itself. Therefore NTs represent themselves the example of the system, which strongly interact with EM-field. Experimental detection of Rabi wave packets confirms the theory, developed in \cite{Slepyan_Yerchak} on the one hand. On the other hand it means, that semiclassical description of spectroscopic transitions in NTs and in the systems like them cannot be appropriate, which seems to be substantially  raising the practical concernment of QED-theory.


\begin{thebibliography}{}
\bibitem{Slepyan} Slepyan G.Ya, Yerchak Y.D, Maksimenko S.A, Hoffmann A, Phys.Lett.A, \textbf{373} (2009) 1374 - 1378
\bibitem{Slepyan_Yerchak} Slepyan G.Ya, Yerchak Y.D, Hoffmann A,  Bass F.G, Phys.Rev.B, 2010
\bibitem{Tavis} Tavis M,  Cummings F W, Phys.Rev., \textbf{170}(2), (1968) 387
\bibitem{Part1} Dovlatova A, Yerchak Y, Yearchuck D, to be submitted
\bibitem{Erchak} Erchak D.P, Efimov V.G, Zaitsev A M, Stelmakh V.F, Penina N.M, Varichenko V S, Tolstych V.P, Nucl.Instr.Meth in Phys.Res. B, \textbf{69} (1992) 443-451  
\bibitem{ErtchakD} Ertchak D.P, Efimov V.G, Stelmakh V.F, J.Applied Spectroscopy, \textbf{64}, N 4 (1997) 433-460
\bibitem{Ertchak} Ertchak D.P, Efimov V.G, Stelmakh V.F, Martinovich V.A, 
Alexandrov A.F, Guseva M B, Penina N.M, Varichenko V S, Karpovich I A, Zaitsev A 
M, Fahrner W R, Fink D, Phys.Stat.Sol.b, \textbf{203} (1997) 529-547
\bibitem{Goldberg} Goldberg I B, Crowe H R, Newman P R, Heeger A.J, MacDiarmid A G, J.Chem.Phys., \textbf{70} (1979) 1132
\bibitem{Kahol} Kahol P K, Mehring M, J.Phys.C, \textbf{19}  (1986) 1045 
\bibitem{Kuroda} Kuroda S, Tokumoto M, Kinoshita N, Shirakawa H, J.Phys.Soc.Jpn,\textbf{51} (1982) 693
\bibitem{Su_Schrieffer_Heeger_1979} Su W-P, Schrieffer J.R, Heeger A.J, Phys.Rev.Lett., \textbf{42}, (1979) 1898 
\bibitem{Su_Schrieffer_Heeger_1980} Su W-P, Schrieffer J.R, Heeger A.J, Phys.Rev.B, \textbf{22} (1980) 2099-2111 
\bibitem{Slater} Slater J C, Nature, \textbf{113} (1924) 307
\bibitem{Lifshitz} Lifshitz E.M, Pitaevsky L.P, Statistical Physics, part 2, M., Nauka, 1978, 448 pp
\bibitem{Heeger_1988} Heeger A.J, Kivelson S, Schrieffer J.R, Su W-P, Rev.Mod.Phys., \textbf{60} (1988) 781-850
\bibitem{Grant} Grant P M, Batra I, Solid State Commun., \textbf{29} (1979) 225
\bibitem{Yearchuck_PL} Yearchuck D, Yerchak Y, Alexandrov A, Phys.Lett.A,  \textbf{373}, N 4 (2009) 489 - 495
\bibitem{Rice} Rice M J, Philpot S R, Bishop A R, Campbell D K, Phys.Rev.B, \textbf{34}, N6 (1986) 4139-4149
\bibitem{Yearchuck_ArXiv} Yearchuck D,  Yerchak E, arXiv: 0709.3382
\bibitem{Leleiter} Leleiter M, Joyes P, J.de Physique \textbf{36} (1975) 343-355
\bibitem{Anglaret} Anglaret E, Bendiab N, Guillard T, Journet C, Flamant G, Laplaze D, Bernier P, Sauvajol J-L, Carbon, \textbf{36}, N12, (1998) 1815-1820
\bibitem{Eklund} Eklund P C, Holden J M, Jishi R A, Carbon, \textbf{33} N7 (1995) 959-972
\bibitem{D_Y} Yearchuck D, Dovlatova A, to be submitted
\bibitem{Tsukanov}  Tsukanov A V, Phys.Rev.B, \textbf{73}, 085308 (2006)
\bibitem{Yearchuck_Doklady} Yearchuck D, Yerchak Y, Red'kov V, Doklady NANB \textbf{51}, N 5 (2007) 57 - 64
\bibitem{Yearchuck_Yerchak} Yearchuck D, Yerchak Y, Kirilenko A, Popechits V, Doklady NANB \textbf{52}, N 1 (2008) 48 - 53
\bibitem{Ertchak_J_Physics_Condensed_Matter} Ertchak D P, Kudryavtsev Yu P, Guseva M B, Alexandrov A F et al, J.Physics: Condensed Matter, \textbf{11}, N3 (1999) 855 -870 
\end{thebibliography}
\end{document}